\documentclass[epj]{svjour}
\usepackage{epsfig}
\usepackage{graphics}
\begin{document}
\title{\boldmath{Quartic Gauge Couplings and the Radiation Zero in pp$\to\mathrm{l}^{\pm}\nu\gamma\gamma$ events at the LHC}}
\author{P. J. Bell\inst{1}
}                     % Do not remove
\institute{ The University of Manchester, Oxford Road, Manchester, M13 9PL, United Kingdom\\paul.bell@cern.ch}
\abstract{
We report a study of the process pp$\to\mathrm{l}^{\pm}\nu\gamma\gamma$ at CERN's Large Hadron Collider, using a leading order partonic-level event generator interfaced to the Pythia program for showering and hadronisation and a with a generic detector simulation.
The process is sensitive to possible anomalous quartic gauge boson couplings of the form WW$\gamma\gamma$. It is shown how unitarity-safe limits may be placed on these anomalous couplings by applying a binned maximum likelihood fit to the distribution of the two-photon invariant mass, $M_{\gamma\gamma}$, below a cutoff of $\sim$1~TeV. Assuming 30~fb$^{-1}$ of integrated luminosity, the expected limits are two orders of magnitude tighter than those available from LEP. It is also demonstrated how the Standard Model radiation zero feature of the q$\bar{\mathrm{q}}\prime\to\mathrm{W}\gamma\gamma$ process may be observed in the difference between the two-photon and charged lepton pseudo-rapidities.
}
\maketitle
\section{Introduction}
\label{sec:intro}
In the Standard Model (SM) the form and strength of the self-interactions of the boson fields are specified by the $\mathrm{SU(2) \times U(1)}$ gauge invariant form of the electroweak sector, through the 
$-\frac{1}{4} {\mathbf W_{\mu\nu}} \cdot {\bf W^{\mu\nu}}$ Lagrangian term.
Any deviations in the self-couplings from their SM expectations may signal the presence of new physics at as yet unprobed energy scales: terms equivalent to anomalous gauge couplings may parametrise the low energy effects of the unknown new physics. 

Whilst both anomalous triple and quartic gauge couplings offer an important test of the
non-Abelian structure of the SM, the anomalous quartic gauge couplings (AQGCs) are also connected to the electroweak symmetry breaking sector; the WWWW quartic coupling must conspire with the Higgs to ensure good high energy behaviour in WW scattering. It has therefore been suggested that the AQGCs may provide a unique window on the mechanism responsible for the symmetry breaking, with any deviations from the SM expected behaviour being a potential sign of some alternative mechanism to that of the Higgs~\cite{bib-godfry,bib-boudjema}.

Possible anomalous triple and quartic gauge couplings may be accessed through di- or tri-vector boson production processes, respectively, but the latter have so far attracted little interest given the modest event rates expected even at the LHC.
The process q$\bar{\mathrm{q}}\prime\to$W$\gamma\gamma$, which is sensitive to the WW$\gamma\gamma$ four-point vertex, represents an obvious starting point to look for a tri-boson signal. Compared to those involving a higher number of heavy bosons, this process requires a relatively low partonic centre of mass energy and gives rise to clean leptonic final states suppressed by the branching of only one massive vector boson. It can be assumed that any deviations  in the couplings  from their SM expectation at the WW$\gamma\gamma$ vertex could be indicative of some general discrepancy in the quartic couplings sector, including the phenomenologically more interesting WWWW case.

Beyond providing a means of studying the AQGCs, W$\gamma\gamma$ production itself is of interest for several other reasons. As in the q$\bar{\mathrm{q}}\prime\to$W$\gamma$ case, the process q$\bar{\mathrm{q}}\prime\to$W$\gamma\gamma$ contains a so called {\emph {radiation zero}} in its amplitude, the observation of which would provide another consistency check of the SM. In addition, W plus two-photon events will need to be considered when making high precision measurements of the W mass.  Finally, W$\gamma\gamma$ production is an irreducible background to the important $\mathrm{H}\to\gamma\gamma$ channel at the LHC. 

The main objective of this work has been to evaluate the expected event rate for  pp$\to\mathrm{l}^{\pm}\nu\gamma\gamma$ ($\mathrm{l}=\mathrm{e},\mu$) at the LHC and investigate the sensitivity of this process to possible anomalous contributions to the WW$\gamma\gamma$ vertex. The theoretical framework for these AQGCs is outlined in section two. The SM expectations from previous Monte Carlo (MC) studies at the Tevatron and LHC are compared in section three, where the adaptation of a generator for use in conjunction with a showering and hadronisation program is also described. 
In the section four, a binned maximum likelihood method is used to place limits on the AQGCs and these expected experimental limits are compared to those obtained from unitarity considerations. The possible observation of the radiation zero is described in the final section.

%%%%%%%%%%%%%%%%%%%%%%%%%%%%%%%%%%%%%%%%%%%%%%%%%%%%%%%%%%%%%%
%%%%%%%%%%%%%%%%%%%%%%%%%%%%%%%%%%%%%%%%%%%%%%%%%%%%%%%%%%%%%%
\section{General formalism for anomalous quartic gauge couplings}
\label{sec:1}

The formalism for possible anomalous terms generating quartic gauge boson self-couplings has been widely discussed 
in the literature~\cite{bib-godfry,bib-boudjema,bib-stirling,bib-boudjema2}. In the parametrisation first 
introduced in~\cite{bib-boudjema}, the two lowest dimension effective Lagrangian terms that give rise to purely quartic couplings involving at least two photons are:
\begin{eqnarray*}
\mathcal{L}_6^0 & = & - \frac{e^2 \beta_0}{16}  F_{\mu\nu}F^{\mu\nu}
\vec{W}^{\alpha} \cdot \vec{W}_{\alpha},
\\
\mathcal{L}_6^{\mathrm{c}} & = & - \frac{e^2 \beta_{\mathrm{c}}}{16}  F_{\mu\alpha}F^{\mu\beta}
\vec{W}^{\alpha} \cdot \vec{W}_{\beta}.
\end{eqnarray*}

These are C and P conserving and are obtained by imposing local U(1)$_{\mathrm{em}}$ gauge symmetry whilst also requiring the global custodial SU(2)$_{\mathrm{c}}$ symmetry that constrains the electroweak parameter $\rho =1$. Noting that the custodial $\mathrm{SU(2)}_{\mathrm{c}}$ field vector is
\[
\vec{W}_{\alpha} = \left( \begin{array}{c}
                           \frac{1}{\surd 2}(W_{\alpha}^+ + W_{\alpha}^-) \\
                           \frac{i}{\surd 2}(W_{\alpha}^+ - W_{\alpha}^-) \\
                           Z_{\alpha} / \cos\theta_W
                           \end{array} \right)
\]
and identifying
\[
\vec{W}_{\alpha} \cdot \vec{W}_{\beta} \to 2(W_{\alpha}^{+} W_{\beta}^{-} + \frac{1}{2\cos^2\theta_W} Z_{\alpha} Z_{\beta}),
\]
then in terms of the physical fields:
\begin{eqnarray*}
\mathcal{L}_6^0  = & - & \frac{e^2 \beta_0^{\mathrm{W}}}{8} F_{\mu\nu}F^{\mu\nu} {W}^{+\alpha}{W}^-_{\alpha}\\
&-& \frac{e^2 \beta_0^{\mathrm{Z}}}{16\cos^2\theta_W}  F_{\mu\nu}F^{\mu\nu} {Z}^{\alpha}{Z}_{\alpha},
\\
\mathcal{L}_6^{\mathrm{c}} = & - & \frac{e^2 \beta_{\mathrm{c}}^{\mathrm{W}}}{16} F_{\mu\alpha}F^{\mu\beta} ({W}^{+\alpha}{W}^-_{\beta}+W^{-\alpha}W^+_{\beta}) \\
&-& \frac{e^2 \beta_{\mathrm{c}}^{\mathrm{Z}}}{16\cos^2\theta_W} F_{\mu\alpha}F^{\mu\beta} {Z}^{\alpha}{Z}_{\beta}.
\label{eqn-lag}
\end{eqnarray*}

\begin{figure}[ht]
\centerline{\epsfig{file=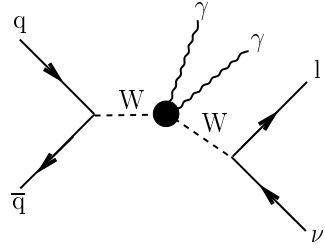, height=3.5cm}}
\caption{The contribution of the WW$\gamma\gamma$ vertex, which may receive an anomalous contribution governed by the coupling parameters  $\beta_0^{\mathrm{W}}$ and $\beta_{\mathrm{c}}^{\mathrm{W}}$, to the process q$\bar{\mathrm{q}}\prime\to\mathrm{l}^{\pm}\nu\gamma\gamma$}
\label{fig:diagram}      
\end{figure}

Thus, both terms generate AQGCs of the form WW$\gamma\gamma$ and ZZ$\gamma\gamma$. The parameters $\beta_0$ and $\beta_{\mathrm{c}}$ are distinguished here for the W and Z vertices to comply with previous experimental measurements in which the couplings were studied independently~\cite{bib-bell}. 
Figure~\ref{fig:diagram} shows how the process  q$\bar{\mathrm{q}}\prime\to\mathrm{l}^{\pm}\nu\gamma\gamma$ includes a contribution from the  WW$\gamma\gamma$ vertex and is thus sensitive to $\beta_0^{\mathrm{W}}$ and $\beta_{\mathrm{c}}^{\mathrm{W}}$.

Through the $F_{\mu\alpha}F^{\mu\beta}$ terms in the effective Lagrangians, the anomalous couplings will scale with the square of the photon energies, so a substantial improvement in the sensitivity can be expected at the LHC over the results from LEP.

%%%%%%%%%%%%%%%%%%%%%%%%%%%%%%%%%%%%%%%%%%%%%%%%%%%%%%%%%%%%%%
%%%%%%%%%%%%%%%%%%%%%%%%%%%%%%%%%%%%%%%%%%%%%%%%%%%%%%%%%%%%%%
\section{\boldmath{Monte Carlo generation of W$\gamma\gamma$ events}}

\subsection{Comparison of programs and published results}
Previous studies have been made of both pp$\to\mathrm{l}\nu\gamma\gamma$ at the LHC by O.~J.~P.~\'Eboli, M.~C.~Gonzalez-Garc\'ia, S.~M.~Lietti and S.~F.~Novaes~\cite{bib-eboli} and p$\bar{\mathrm{p}}\to\mathrm{l}\nu\gamma\gamma$ at the Tevatron by U.~Baur, T.~Han, N.~Kauer, R.~Sobey and D.~Zeppenfeld~\cite{bib-baur}. The MC program used in these works have been obtained from the corresponding authors, and are referred to here as the Lietti and Baur MCs, respectively; they are described fully in the corresponding publications.
Both programs are based on Madgraph-generated amplitudes~\cite{bib-mgraph} that take into account all leading order diagrams for the $\mathrm{l}^\pm\nu\gamma\gamma$ final state. Finite W width effects are included and all partons are assumed to be massless. Both programs produce weighted events, the Lietti MC relying on Vegas~\cite{bib-vegas} for the phase space integration and the Baur code making use of a custom three body phase space generator. The important difference between the programs is that whereas the Baur MC generates only SM events, the Lietti code includes the AQGC contribution to the WW$\gamma\gamma$ vertex, parametrised by the $\beta_0$ and $\beta_{\mathrm{c}}$ parameters. 

Whilst the Lietti program forms the basis of the work reported here, it is prudent to first compare the SM expectations from the two generators in order to validate the programs and our usage of them.
The generator-level cuts applied in order to approximately simulate the detector acceptance in the previous studies are summarised in table~\ref{tab:cuts}. For the Tevatron (Baur MC), only the $\mathrm{W}^-\to\mathrm{e}^- \nu$ channel was considered, the  W$^+$ channel not being implemented in the MC. At the LHC (Lietti MC) the complete $\mathrm{W}^\pm\to\mathrm{l}^{\pm} \nu$ ($\mathrm{l}=\mathrm{e},\mu$) final state was studied. 
The MRS~(A) and  MRS~(G) sets of proton structure functions were used for the Tevatron and LHC studies, respectively, with the factorisation scale in both cases being set equal to the parton centre of mass energy.

The prediction from each program ``as provided'' was first verified against the corresponding published result. The Baur MC faithfully reproduced the reported cross-section for p$\bar\mathrm{p}\to \mathrm{e}^{-}\nu\gamma\gamma$ at the Tevatron, and likewise the Lietti MC for pp$\to \mathrm{l}^\pm\nu\gamma\gamma$ at the LHC (see table~\ref{tab:res}).
\begin{centering}
\begin{table*}[ht]
\caption{The selections applied in~\cite{bib-eboli,bib-baur} on the previously studied channels for W$\gamma\gamma$ production at the Tevatron and LHC. Cuts were applied on the photon transverse momenta, $p_T^{\gamma}$, the charged lepton transverse momenta, $p_T^{\mathrm{l}}$ and pseudo-rapidity, $|\eta_{\mathrm{l}}|$ and on the photon and charged lepton separations, $\Delta R$. In addition, in order to suppress photon radiation from the final state charged lepton, the transverse mass of the $(\mathrm{l},\nu)$ system, $M_T(\mathrm{l},\nu)$, was required to satisfy  $M_T(\mathrm{l},\nu)>70$~GeV  at the LHC and  $65<M_T(\mathrm{l},\nu)<100$~GeV at the Tevatron.  No energy smearing or efficiencies are applied.}
\label{tab:cuts}
\begin{tabular}{|l|l|c|c|c|c|c|c|c|}
\hline
~~Collider~~ & ~~Process~~ &  ~~$p_T^{\mathrm{e}}$~[GeV]~~ & ~~$p_T^{\mu}$~[GeV]~~ & ~~$p_T^{\gamma}$~[GeV]~~ & $|\eta_{\mathrm{e}}|$ &  $|\eta_{\mu}|$ & $\Delta R_{\gamma\mathrm{l}}$ & $\Delta R_{\gamma\gamma}$   \\
\hline
~~Tevatron & ~~p$\bar\mathrm{p}\to \mathrm{e}^{-}\nu\gamma\gamma$  &  $>$15  &  n/a      &  $>$15  &  ~~~$<$2.5~~~  &   ~~~n/a~~~     &  ~~~$>$0.7~~~  &  ~~~$>$0.4~~~    \\
~~LHC & ~~pp$\to \mathrm{l}^\pm\nu\gamma\gamma$  & $>$20  &  $>$25  &  $>$20  &  ~~~$<$2.5~~~  &  ~~~$<$1.0~~~  &  ~~~$>$0.4~~~  &  ~~~$>$0.4~~~     \\
\hline
\end{tabular}
\end{table*}
\end{centering}
\begin{centering}
\begin{table*}[ht]
\caption{Comparison of the previously published~\cite{bib-eboli,bib-baur} expected cross-sections for W$\gamma\gamma$ production at the Tevatron and the LHC with the results obtained from the Baur and Lietti MCs. The selections at each collider are as defined in table~\ref{tab:cuts}.}
\label{tab:res}
\begin{tabular}{|l|l|l|l|}
\hline
~~~Collider~~~  & ~~~~~~Process~~~~~~ & ~~~MC & ~~~Cross-section (fb)~~~  \\
\hline
~~~Tevatron        &  ~~~~~~$\mathrm{p}\bar\mathrm{p}\to \mathrm{e}^{-}\nu\gamma\gamma$~~~~~~ & ~~~Baur (published result)     &  ~~~0.50 \\
                &                                                     & ~~~Baur (from MC provided)          &  ~~~0.50 \\
\hline
~~~LHC             &  ~~~~~~$\mathrm{pp}\to \mathrm{l}^\pm\nu\gamma\gamma$~~~~~     & ~~~Lietti (published result)    & ~~~1.76 \\
                &                                                     & ~~~Lietti (from MC provided)         & ~~~1.79 \\
\hline
~~~LHC             &  ~~~~~~$\mathrm{pp}\to \mathrm{e}^-\nu\gamma\gamma$~~~~~~       & ~~~Lietti (from MC provided)         & ~~~0.546 \\
                &                                                     & ~~~Baur  (from MC modified for LHC)~~~    & ~~~0.672 \\
                &                                                     & ~~~Lietti (from MC corrected)           & ~~~0.675 \\
\hline
\end{tabular}
\end{table*}
\end{centering}
However, when the Baur MC was modified to generate pp$\to \mathrm{e}^-\nu\gamma\gamma$ events at the LHC, the expected cross-section was found to be about 25\%  higher than that obtained from the Lietti program. Conversely, a similar modification of the Lietti MC for the generation of $\mathrm{p}\bar{\mathrm{p}}$ collisions at the Tevatron gave results comparable to those published.
This discrepancy was due to the inclusion of only the $\mathrm{u}\bar{\mathrm{d}} \to \mathrm{l}^+\nu\gamma\gamma$ and $\mathrm{d}\bar{\mathrm{u}} \to \mathrm{l}^-\nu\gamma\gamma$  contributions to the total cross-section within the Lietti MC. We have made a ``corrected'' version of this program by including the missing valence-sea quark terms, which make a significant contribution to the total cross-section at the LHC but play only a small role at the Tevatron.

\subsection{Adaptation of a parton-level Monte Carlo for use with a showering and hadronisation generator}

We have adapted the corrected Lietti MC for use with a showering and hadronisation generator (SHG) by adding a routine to write out events of unit weight in the Les Houches format~\cite{bib-lesh}. This unweighting routine works in the usual way, with each event being selected with a probability $P=w/w_{\mathrm{max}}$, where $w$ is the weight of the event and $w_{\mathrm{max}}$ is some maximum weight found from a large sample of events.
In order to be able to label the incoming quarks in the event record, the weight $w$ is expressed as the sum of the contributions from the different initial state  $\mathrm{q}\bar{\mathrm{q}}$ combinations, in proportion to the parton distributions in that event, i.e. $w= w_{\mathrm{d}\bar{\mathrm{u}}} + w_{\mathrm{u}\bar{\mathrm{d}}} + ...$\,. 
For each unweighted event the initial state can therefore be assigned on a statistical basis and the incoming partons labelled accordingly.

\begin{figure*}[ht]
\centerline{\epsfig{file=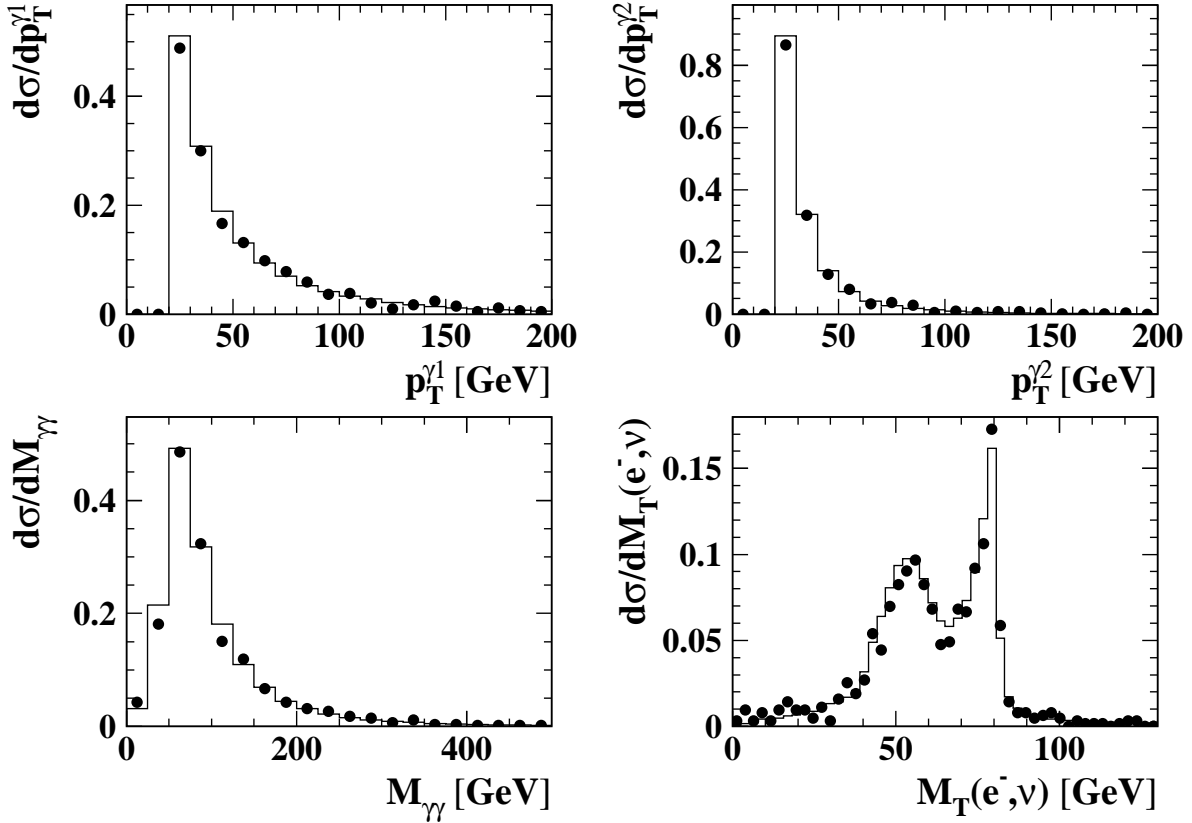, height=11cm}}
\caption{Distributions of the transverse momenta of the two photons, $p_T^{\gamma_1}$ and $p_T^{\gamma_2}$ (where $p_T^{\gamma_1} > p_T^{\gamma_2}$), the invariant mass of the two-photon pair, $M_{\gamma\gamma}$ and the transverse mass of the electron and neutrino, $M_T(\mathrm{e},\nu)$ for pp$\to \mathrm{e}^-\nu\gamma\gamma$ events at the LHC. The solid lines use weighted events from the Baur MC (which is for the W$^-$ channel only) whereas the points use unit-weight events selected from  the Lietti MC pp$\to \mathrm{l}^\pm\nu\gamma\gamma$ event record by requiring that the incoming $\mathrm{q}\bar{\mathrm{q}}$ mixture be consistent with W$^-$ production.}
\label{fig-unwgt}
\end{figure*}

We illustrate the correct functioning of the unweighting routine in figure~\ref{fig-unwgt}.
As mentioned above, the Baur MC generates only W$^-$ events, meaning that the input $\mathrm{q}\bar{\mathrm{q}}$ mixture is always a combination of a down-type quark with an anti-up-type quark. Various distributions have been made using events with only these initial states selected from the full Lietti MC unweighted event record. In the figure these distributions are superimposed on those obtained directly from the Baur MC weighted events. The distributions are scaled according to the selected cross-sections from the two MCs. The result not only confirms the good agreement between the Baur and corrected Lietti programs, as already seen in table 2, but also demonstrates the correct unweighting and labelling of events from the Lietti MC. 

The adapted, corrected Lietti event generator, which we now name W2PHO~\cite{bib-w2pho}, is available for download from the HepForge~\cite{bib-hepforge} website. This program was then employed in the subsequent work described here.

\section{\boldmath{Studying the WW$\gamma\gamma$ anomalous coupling in pp~$\to\mathrm{l}^\pm\nu\gamma\gamma$ at the LHC}}

\subsection{Generation and simulation of signal events}

The W2PHO program was used to generate samples of  pp~$\to\mathrm{l}^\pm\nu\gamma\gamma$ ($\mathrm{l}=\mathrm{e},\nu$) events at the LHC centre of mass energy of $\sqrt{s}=14$~TeV. 
The CTEQ5L structure functions were used. The Les Houches format output files were provided as input to Pythia version 6.4~\cite{bib-pythia} to perform the showering and hadronisation. Subsequently, the response of a generic LHC detector was simulated using the PGS program~\cite{bib-pgs} employing the default LHC detector description parameter set. A minimal set of cuts were applied on the reconstructed quantities to give the most inclusive selection within a likely trigger acceptance:
\begin{itemize}
\item
Transverse momenta of both photons, $p_T^{\gamma}>15$~GeV
\item
Transverse momentum charged lepton, $p_T^{\mathrm{l}}>25$~GeV
\item
Missing transverse momentum, $p_T^{\mathrm{miss}}>25$~GeV
\item
Pseudo-rapidity, $\eta$, of all charged leptons and photons satisfying $|\eta|<2.5$
\end{itemize}
Within these cuts an expected reconstructed cross-section of 8.6~fb was obtained, corresponding to around 260 SM events in 30~fb$^{-1}$ of integrated luminosity. It should be noted that the cross-section increases approximately logarithmically as the cut on the photon transverse momenta is reduced.

\subsection{Observing and constraining the anomalous quartic gauge couplings}

\begin{figure*}[ht]
\centerline{\epsfig{file=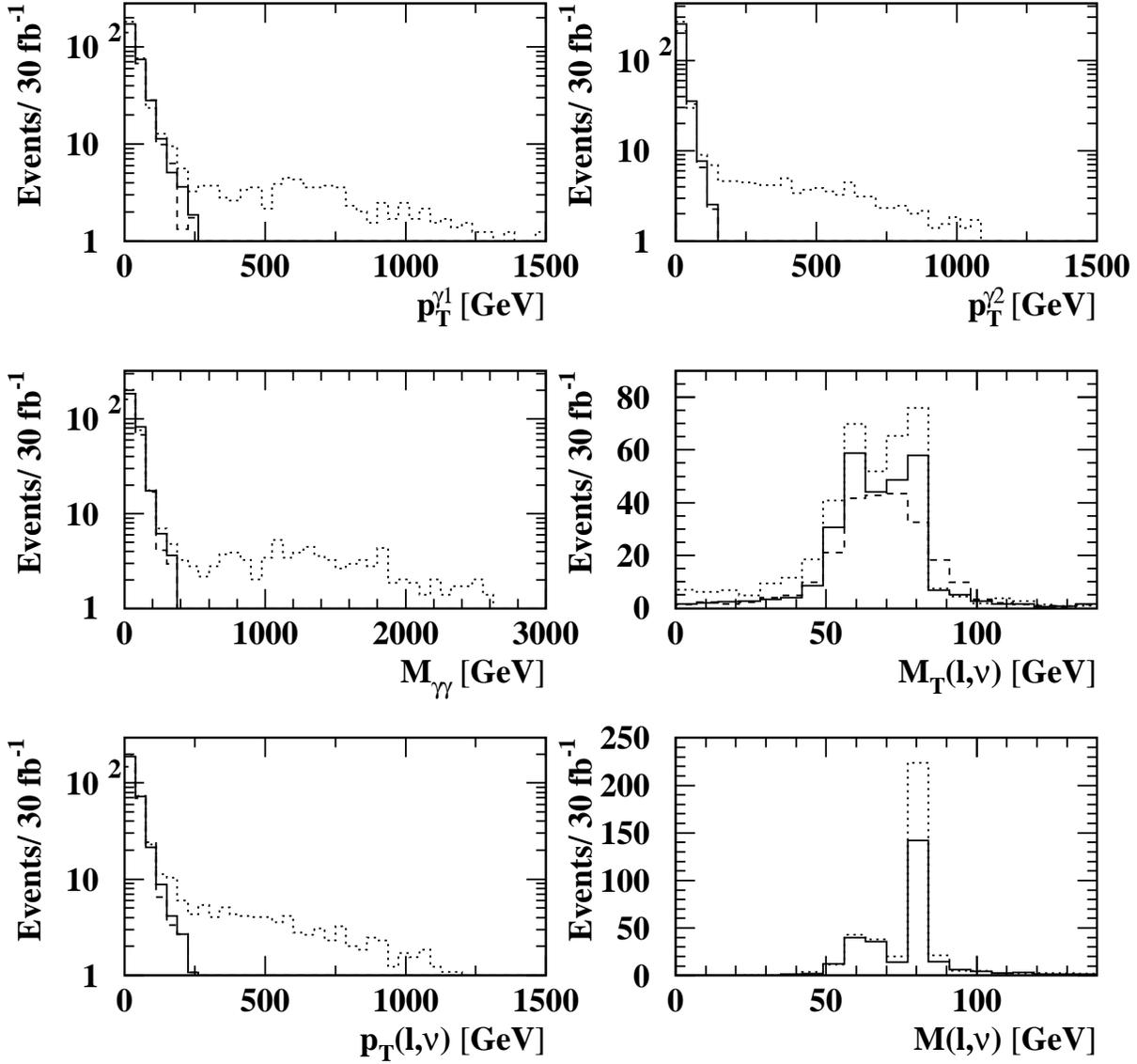, height=16.5cm}}
\caption{ Distributions of the transverse momenta of the two photons, $p_T^{\gamma_1}$ and $p_T^{\gamma_2}$ (where $p_T^{\gamma_1} > p_T^{\gamma_2}$), the invariant mass of the two-photon system, $M_{\gamma\gamma}$,  the transverse mass of the  $(\mathrm{l},\nu)$ system, $M_T(\mathrm{l},\nu)$, the transverse momentum of the $(\mathrm{l},\nu)$ system, $p_T(\mathrm{l},\nu)$ and the invariant mass of the $(\mathrm{l},\nu)$ system, $M(\mathrm{l},\nu)$ . The solid lines show the generator-level SM expectation, the dashed line is for the reconstructed events after processing in PGS (hence $M(\mathrm{l},\nu)$ is not available). The dotted line represents the generator level expectation for $\beta_0=0.0001$~GeV$^{-2}$, which is about 0.02 of the LEP limit.}
\label{fig-6plots}
\end{figure*}

 Since the effective Lagrangian terms for the AQGCs are linear in the coupling constants $\beta_0$ and $\beta_{\mathrm{c}}$ the total cross-section gains a quadratic dependence on each parameter. 
A simple counting method could therefore be employed to compare the total number of events observed with the number expected parametrised as a function of $\beta_0$ and $\beta_{\mathrm{c}}$. Much greater sensitivity can be obtained, however, by making use of the effect of any AQGC contribution on the shapes of various distributions, as shown in figure~\ref{fig-6plots}. 

Of particular interest in figure~\ref{fig-6plots} are the distributions of the lepton-neutrino invariant mass, $M(\mathrm{l},\nu)$ and  lepton-neutrino transverse mass, $M_T(\mathrm{l},\nu)$. The AQGCs clearly contribute to the $M(\mathrm{l},\nu)$ distribution exclusively at the W mass peak, which is to be expected given the diagram of figure~\ref{fig:diagram}. Events lying below this peak arise due to the final state photon radiation from the charged lepton pulling the $M(\mathrm{l},\nu)$ mass down below that of the W: these events are not part of the pp~$\to\mathrm{W}(\to\mathrm{l}\nu)\gamma\gamma$ contribution to the total cross-section where the sensitivity to the WW$\gamma\gamma$ vertex lies.
Whilst the $M(\mathrm{l},\nu)$ distribution cannot be obtained from experimental data the transverse mass distribution, $M_T(\mathrm{l},\nu)$ can be used as an alternative: in the SM this distribution is also peaked at $M_{\mathrm{W}}$ for events in which neither photon is emitted from the charged lepton. The usual approach to isolate the W$(\to \mathrm{l}\nu)\gamma\gamma$ part of total cross-section is therefore to cut away the region of  $M_T(\mathrm{l},\nu)$ below, for example, 70~GeV, as advocated in~\cite{bib-eboli}. In figure~\ref{fig-6plots}, however, it can be seen in the $M_T(\mathrm{l},\nu)$ distribution that the effect of the AQGC is not confined to the region around $M_{\mathrm{W}}$. The reason lies in the distribution of the transverse momentum of the (l,$\nu$) system, $p_T(\mathrm{l},\nu)$: in the AQGC scenario, the system is boosted in the transverse direction, which in turn distorts the shape of the  $M_T(\mathrm{l},\nu)$ distribution and increases the number of events in the region below $M_{\mathrm{W}}$~\cite{bib-wtrans}. As a result, cutting on  $M_T(\mathrm{l},\nu)$ below the W mass gives an overall reduction in the sensitivity to any  AQGC, and we do not apply such a cut here.
%(cite http://prola.aps.org/pdf/PRL/v50/i22/p1738_1)

The bulk of the SM contribution lies in the region of low photon transverse momenta, where the anomalous effects are also small, which could be cleanly cut away. However, whilst this would greatly improve the sensitivity of a counting method to the influence of the any AQGC, the method would remain critically dependent on an accurate prediction of the overall rate, on which there are many experimental (e.g. luminosity measurement) and theoretical (e.g. next to leading order effects) uncertainties. 
In this study we have applied a binned maximum likelihood method to the entire shape of various distributions, without discarding any of the SM data. The distribution most sensitive to any AQGC was identified by comparing the expected widths of the 95\% confidence level intervals given a perfectly SM-like observation with 30~fb$^{-1}$ of integrated luminosity.  We found that the invariant mass of the two-photon system, $M_{\gamma\gamma}$, offered the optimal sensitivity. The statistical sensitivity achievable from the binned fit to the full $M_{\gamma\gamma}$ distribution is higher than that obtained from a counting method in which tight cuts are first applied to remove the SM background.

\subsection{Simulation of background contribution}

The dominant backgrounds are anticipated to be W$\gamma$+jets and W+jets events in which one or two jets are mis-identified as photons.
The probability for a jet to be mis-identified in this way is given by 1/$R_{\mathrm{jet}}$ where $R_{\mathrm{jet}}$ is referred to as the jet rejection factor and is a property of the detector performance and reconstruction software. Since the cross sections for the background processes  are several order of magnitudes higher than that of the signal process a high jet rejection factor is required if the background is not to dominate.

The expected background contribution arising from mis-identified jets has been evaluated using Alpgen~\cite{bib-alpgen} samples of W$\gamma$+1jet and W+2jet events. The generated events were showered and hadronised in Pythia 6.4, employing the MLM parton matching scheme~\cite{bib-mlm}, and simulated in PGS. For the  W$\gamma$+1jet events, an event was selected with weight 1/$R_{\mathrm{jet}}$ for any jet which, if relabelled as a photon, allowed the event to pass the signal selection cuts. For the  W+2jet events, an event was selected with weight 1/$R_{\mathrm{jet}}^2$ for any pair of jets which, if relabelled as photons, allowed the event to pass the signal selection cuts. 

\begin{figure*}[hbt]
\centerline{\epsfig{file=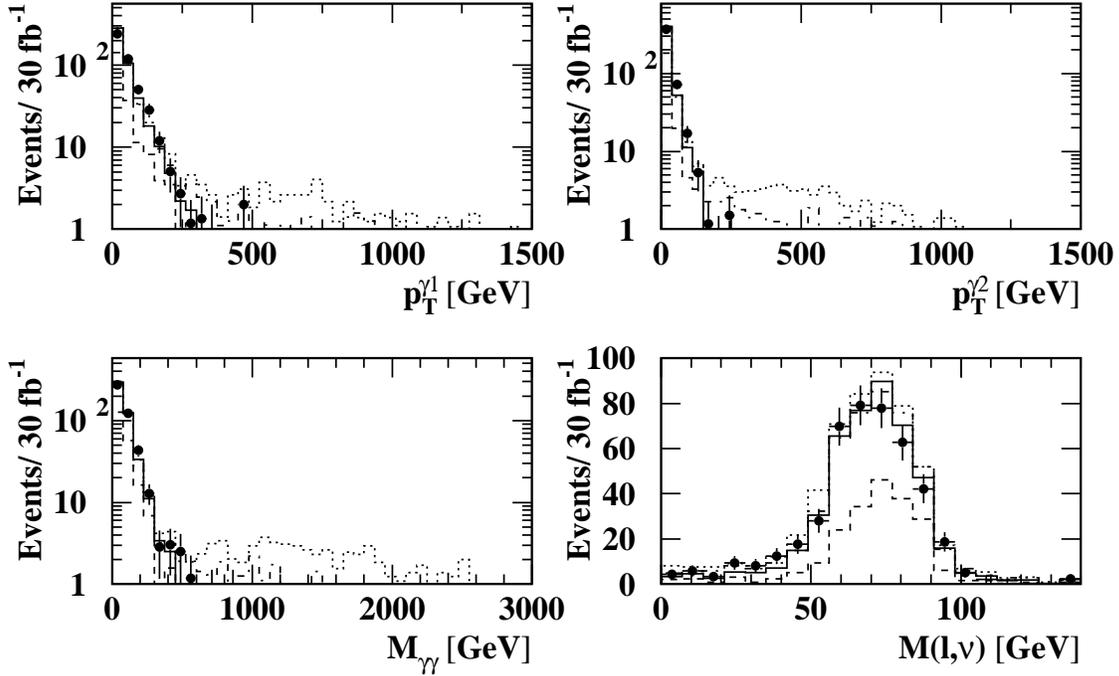, height=10.6cm}}
\caption{ Distributions of the transverse momenta of the two photons, $p_T^{\gamma_1}$ and $p_T^{\gamma_2}$ (where $p_T^{\gamma_1} > p_T^{\gamma_2}$), the invariant mass of the two-photon system, $M_{\gamma\gamma}$ and the transverse mass of the  $(\mathrm{l},\nu)$ system, $M_T(\mathrm{l},\nu)$ assuming 30~fb$^{-1}$ of integrated luminosity. The solid lines show the expected reconstructed SM signal plus background expectation. Of this, the dashed line shows the total background contribution from W$\gamma$+jets and W+jets events assuming a jet rejection factor of 2000. The dotted lines and dashed-dotted lines are the expected reconstructed signal plus background expectation for $\beta_0=0.0001$~GeV$^{-2}$ and  $\beta_c=0.0001$~GeV$^{-2}$, respectively. The points are for one LHC ``experiment'' according to the SM. }
\label{fig-4plots}
\end{figure*}

The results are shown in figure~\ref{fig-4plots}, which presents the total signal plus background expectation assuming a jet rejection factor of 2000. It can be seen that the background contribution lies in the region populated by the SM signal process and is well separated from any possible AQGC signal.

\subsection{Possible experimental limits on the anomalous couplings}

To place confidence limits on the anomalous coupling parameters 10\,000 samples of events were generated, the number of events in each sample being obtained from a Poisson distribution with a mean equal to the SM signal plus background expectation in 30fb$^{-1}$ of data. The results from one such ``experiment'' are shown superimposed in the plots of figure~\ref{fig-4plots}. 
For each sample, the 1-dimensional log-likelihood curves for $\beta_0$ and $\beta_{\mathrm{c}}$ were evaluated and from these the 95\% confidence level limits on the parameters found. The couplings were varied independently with the parameter not under test fixed at its SM value (zero). These limits were then averaged over the 10\,000 experiments to give the final results, which are presented in table~\ref{tab:lims}. 

With 30fb$^{-1}$ of integrated luminosity, the limits are more than two orders of magnitude tighter than those available from LEP (OPAL). The limits obtained assuming 10 and 100~fb$^{-1}$ of data are also shown for comparison. 

\begin{table}[h]
\caption{The expected 95\% confidence level limits on the coupling parameters $\beta_0$ and $\beta_{\mathrm{c}}$ assuming 10, 30 and 100~fb$^{-1}$ of integrated luminosity and assuming a jet rejection factor of 2000.}
\label{tab:lims}  
\begin{tabular}{|l|c|c|}
\hline
              & $\beta_0$                     &  $\beta_{\mathrm{c}}$ \\
\hline
~10~fb$^{-1}$~  & ~$(-2.98, 3.28)\times 10^{-5}$ & ~$(-5.00, 4.92)\times 10^{-5}$ \\
~30~fb$^{-1}$~  & ~$(-1.85, 2.19)\times 10^{-5}$ & ~$(-3.19, 3.21)\times 10^{-5}$ \\
~100~fb$^{-1}$~ & ~$(-1.16, 1.50)\times 10^{-5}$ & ~$(-2.03, 2.14)\times 10^{-5}$ \\
\hline
\end{tabular}
\end{table}

\begin{figure}[ht]
\centerline{\epsfig{file=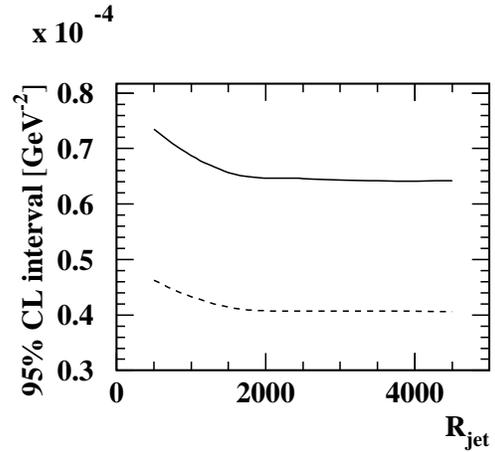, height=6.0cm}}
\caption{The widths of the 95\% confidence level intervals for $\beta_0$ (solid line) and $\beta_{\mathrm{c}}$ (dashed line) for 30~fb$^{-1}$ of integrated luminosity as a function of the jet rejection factor.}
\label{fig:jr}      
\end{figure}

To understand the effects of the background from the mis-identified jets, the confidence level limits for the 30~fb$^{-1}$ case were also found as a function of the jet rejection factor. The results are shown in figure~\ref{fig:jr} where it can be seen that increasing the performance beyond 1000 does not significantly improve the limits. 

\subsection{Comparison to unitarity constraints}

The effective Lagrangian terms generating the AQGCs spoil the gauge structure of the model, which can lead to unitarity violation at relatively low energies. 
To preserve unitarity up to higher energy scales, the conventional procedure is to modify the bare coupling parameters with an energy dependent form factor. A typical choice is the generalised dipole form factor, which in this case would be applied as
\begin{equation}
\beta \rightarrow \left( 1+ \frac{M_{\gamma\gamma}^2}{\Lambda_{\mathrm{FF}}^2} \right)^{-n} \times \beta.
\label{eqn-ff}
\end{equation}
For values of $M_{\gamma\gamma}$ above the form factor scale, $\Lambda_{\mathrm{FF}}$, this has the effect of pushing the AQGCs back down towards the SM prediction. The strength of this effect depends on the choice of $n$. For large $n$ the form factor is effectively a cutoff on the effects of the anomalous couplings at $\Lambda_{\mathrm{FF}}$, so that for $M_{\gamma\gamma}>\Lambda_{\mathrm{FF}}$ any distribution becomes constrained to its SM expectation. 
The undesirable consequence of applying such a form factor is that any limits found will depend on the choices of $n$ and $\Lambda_{\mathrm{FF}}$.  

\begin{figure*}[ht]
\centerline{\epsfig{file=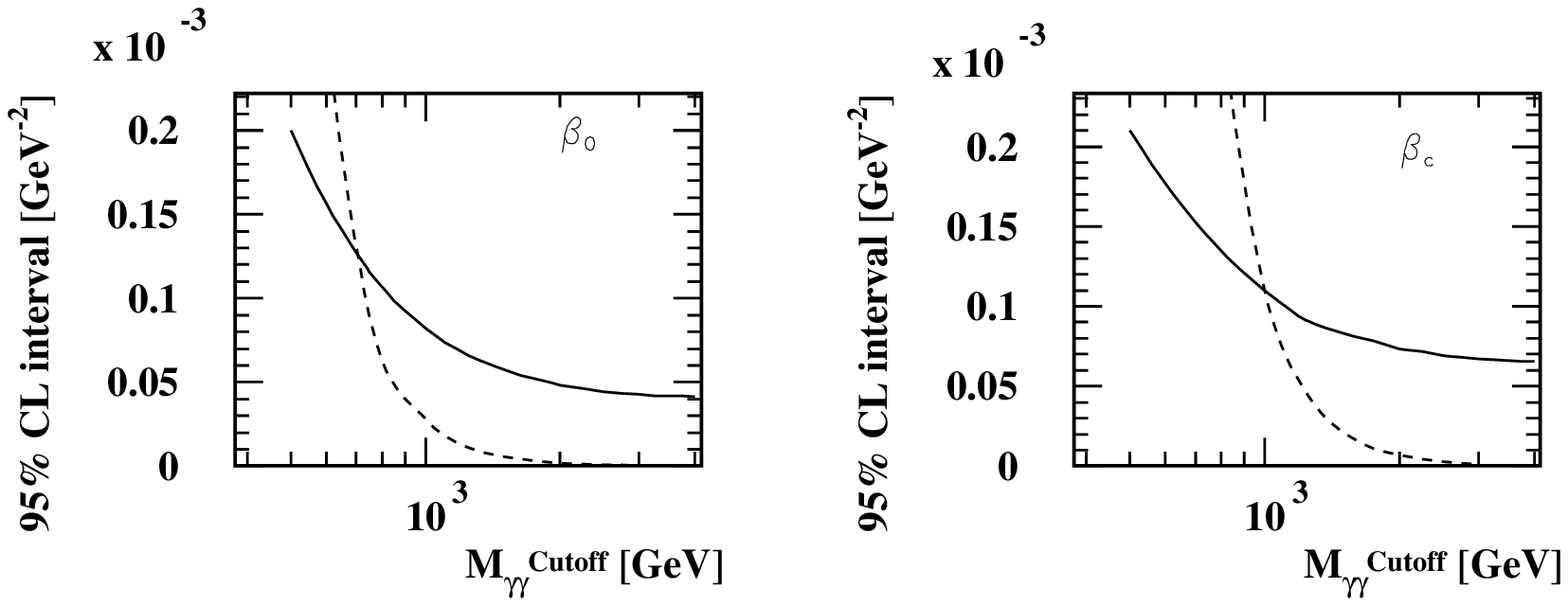, height=6.3cm}}
\caption{ The widths of the 95\% confidence level intervals (solid lines) for $\beta_0$ and $\beta_{\mathrm{c}}$ as a function of the cutoff applied to the $M_{\gamma\gamma}$ invariant mass assuming 30~fb$^{-1}$ of integrated luminosity. Superimposed (dotted) are the unitarity limits from equations~4.3 and~4.4. In each plot the region to the right of the dotted line is excluded by unitarity and above the solid lines by experiment.}
\label{fig-lims}
\end{figure*}

The limits so far found refer to the bare couplings and cannot be assumed to be unitarity-safe. Rather than guaranteeing this by including some arbitrary, energy dependent form factor correction, 
an alternative approach is taken, as advocated in~\cite{bib-dobbs}, whereby the limits are evaluated as a function of a cutoff applied to the mass scale $M_{\gamma\gamma}$. 
This is plotted in figure~\ref{fig-lims}: the experimental 95\% confidence level intervals found for $M_{\gamma\gamma}^{\mathrm{cutoff}}=1$~TeV, for example, use only events for which   $M_{\gamma\gamma}$ falls below this value.
As the cutoff is increased, the experimental limits turn asymptotic, tending towards the values in table~\ref{tab:lims}: they do so at around 3~TeV, which is recognised as the ultimate reach of the experiment on the $M_{\gamma\gamma}$ scale. 

To determine if the asymptotic limits are unitarity-safe, they can be compared to 
the unitarity constraints derived from the 2~$\to$~2 inelastic scattering process $\gamma_1\gamma_2\to \mathrm{W}_1 \mathrm{W}_2$. 
The tightest energy dependent constraint arises from the J~$=0$ partial wave which can be written as~\cite{bib-eboli}
\begin{equation}
\left( \frac{\alpha \beta M_{\gamma\gamma}^2}{16} \right)^2 \left(1-\frac{4{M_{\mathrm{W}}^2}}{M_{\gamma\gamma}^2}\right)^\frac{1}{2} \left(3-\frac{M_{\gamma\gamma}^2}{M_{\mathrm{W}}^2}+\frac{M_{\gamma\gamma}^4}{4 M_{\mathrm{W}}^4}\right) \leq  N,
\end{equation}
where $M_{\gamma\gamma}$ is the invariant mass of the two photons and $N=\frac{1}{4}$ for $\beta = \beta_0$ and $N=4$ for $\beta = \beta_{\mathrm{c}}$. 
%These inequalities can be interpreted as a unitarity constraint on the coupling $\beta$ at an arbitrary $\gamma\gamma$ mass scale. 
Defining $\Lambda_{\gamma\gamma}$ as the two-photon invariant mass scale at which unitarity is violated, for $\Lambda_{\gamma\gamma}\gg M_{\mathrm{W}}$ the inequalities yield the following constant unitarity constraints on the couplings:
\begin{eqnarray}
\beta_{0} & \leq & \frac{13.1\mathrm{TeV}^2}{\Lambda_{\gamma\gamma}^4}~\mathrm{TeV}^{-2},
\\
\beta_{\mathrm{c}} & \leq & \frac{52.4\mathrm{TeV}^2}{\Lambda_{\gamma\gamma}^4}~\mathrm{TeV}^{-2}.
\label{eqn-limits1}
\end{eqnarray}
These inequalities are shown superimposed in figure~\ref{fig-lims}. In the region to the right of these dotted lines any new physics cannot be described by the effective Lagrangian theory.
Assuming 30~fb$^{-1}$ of integrated luminosity, the experimental limits at the reach of the machine lie well inside this region, and thus are weaker than those imposed by unitarity. The experiment will give tighter constraints on the couplings only up to $M_{\gamma\gamma}\sim 750$~GeV for $\beta_0$ and $M_{\gamma\gamma}\sim 1000$~GeV for $\beta_{\mathrm{c}}$. \\
\indent To obtain unitarity-safe limits, a form factor like (\ref{eqn-ff}) could be 
applied to the couplings with the scale $\Lambda_{\mathrm{FF}}$ set to 750~GeV and 
1000~GeV for $\beta_0$ and $\beta_{\mathrm{c}}$, respectively. The limits as a 
function of  $M_{\gamma\gamma}^{\mathrm{cutoff}}$ would then turn asymptotic around 
these values, i.e. within the unitarity-allowed regions of the plots of
figure~\ref{fig-lims}. However, we have already noted that the effect of the form 
factor is to constrain the AQGC contribution to the SM prediction in the region 
where $M_{\gamma\gamma}>\Lambda_{\mathrm{FF}}$. It would therefore make no sense to 
use any data collected in this region to measure the AQGCs; any effects would be 
highly overestimated in overcoming the suppression of the form factors. Consequently, 
it has been argued that the scale chosen for $\Lambda_{\mathrm{FF}}$ should not be 
within the reach of the experiment~\cite{bib-dobbs}. Applying a generalised dipole 
form factor like (\ref{eqn-ff}) is therefore an unsuitable way to ensure unitarity here, 
and instead limits on the bare couplings should be found using a restricted range of 
the $M_{\gamma\gamma}$ distribution. 

It can be seen in figure~\ref{fig-lims} that the 
experimental limits at the edge of the unitarity-allowed region of the 
$(M_{\gamma\gamma},\beta)$ plane are weakened by approximately a factor two compared 
to the asymptotic limits, and so will remain around two orders of magnitude stricter 
than those available from LEP (OPAL).

\section{\boldmath{Observation of the radiation zero in W$\gamma\gamma$ events at the LHC}}

At the Born level in the SM the amplitude for $\mathrm{q}\overline{\mathrm{q}}'\to \mathrm{W^{\pm}}\gamma\gamma$  exhibits a cancellation for $\cos \theta ^{\star} = \mp\frac{1}{3}$ when the two photons are collinear, where $\theta ^{\star}$ is the angle between the incoming quark and the W boson in the parton centre of mass frame~\cite{bib-r0wgg6}. 
It has been shown that this so called {\emph {radiation zero}} only gradually vanishes as the opening angle of the photons is increased, and may be observed experimentally as a ``dip'' in the distribution of $\Delta \eta = \eta_{\gamma\gamma} - \eta_{\mathrm{l}}$, where $\eta_{\gamma\gamma}$ and $\eta_{\mathrm{l}}$ are the pseudo-rapidities of the two-photon system and the charged lepton, respectively~\cite{bib-baur}.

\begin{figure}[ht]
\centerline{\epsfig{file=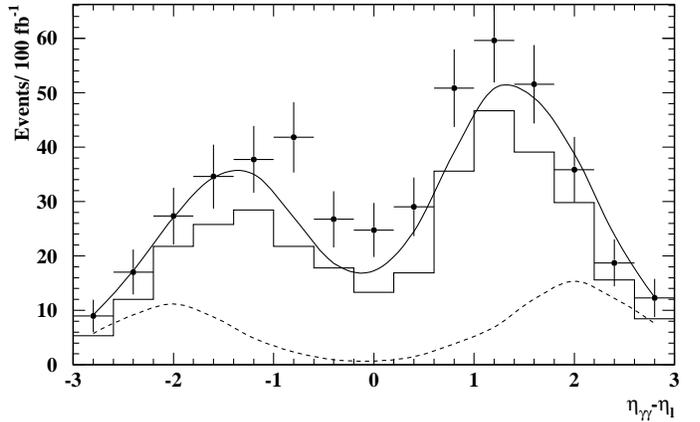, height=6.3cm}}
\caption{The pseudo-rapidity separation of the charged lepton from the two-photon system for W$\gamma\gamma$ production at the LHC, showing the radiation zero ``dip''. The kinematic cuts are as described in section~4.1, with the cut on the charged lepton and photon separations hardened to $\Delta R > 0.7$ and an additional requirement imposed on the $(\mathrm{l},\nu)$ transverse mass of $M_T(\mathrm{l},\nu) > 70$~GeV. The solid curve shows the generator-level expectation for the ``signed distribution'', the histogram is the signal after simulation in PGS and the points are the signal plus background assuming a jet rejection factor of 4000. The dotted line is the generator-level expectation after additionally requiring that the opening angle of the two photons in the W$\gamma\gamma$ centre of mass frame satisfies $\cos(\theta_{\gamma\gamma}^*)>0$.}
\label{fig:r0}
\end{figure}

At a pp collider such as the LHC, the symmetric beams mask the asymmetry of the radiation zero and the dip in $\Delta \eta$ occurs at zero. However, it has been suggested~\cite{bib-dobbs} that the distribution may be ``signed'' according to the longitudinal direction of the final state system. Since the quark is statistically most likely to come from the valence distribution in the proton, whereas the anti-quark has to come from the sea distribution, the quark will tend to carry a larger momentum fraction than the anti-quark and the   $\mathrm{l}\gamma\gamma$ system will most likely be boosted in the quark direction. The longitudinal direction of the $\mathrm{l}\gamma\gamma$ system can therefore be evaluated and if found to be in the backward direction the sign of the $\Delta \eta$ distribution reversed. To take account of the inherent  sign difference between the W$^+$ and W$^-$ cases the sign of $\Delta \eta$ is also reversed for the W$^-$ events. This signing maintains the asymmetry of the radiation zero, as can be seen in figure~\ref{fig:r0}. 

For the plot in  figure~\ref{fig:r0} the event selection of section~4.1 has been tightened by imposing a cut on the charged lepton and photon separations of $\Delta R > 0.7$ and on the $(\mathrm{l},\nu)$ transverse mass of $M_T(\mathrm{l},\nu) > 70$~GeV. These cuts reduce the effects of photon radiation in the directions of the final state charged lepton and initial state quarks, respectively, which would otherwise obscure the radiation zero. The background contribution from the mis-identified jets also acts to fills in the dip, and a tighter jet rejection factor of 4000 is therefore used. With these additional constraints the dip is clearly visible in 100~fb$^{-1}$ integrated luminosity. 

Since the exact cancellation requires that the two photons be collinear, it has been suggested~\cite{bib-baur} that the radiation zero can be enhanced by  cutting on the two-photon opening angle, $\cos(\theta_{\gamma\gamma})$. Contrary to what was reported in~\cite{bib-baur}, however, it was found here that such a cut is effective only if applied in the centre of mass system. 
Boosting to the centre of mass frame requires the knowledge of the missing longitudinal momentum, which can be reconstructed with a two-fold ambiguity if it is assumed that the missing transverse momentum belongs exclusively to the neutrino and that the W is produced on-shell.  Experimentally it is not possible to determine which of the two solutions for the missing longitudinal momentum is the correct one, but statistically it is most likely to be the one which gives the smallest mass to the W$\gamma\gamma$ system. Using this ``minimum mass solution'', the events can be boosted to the centre of mass frame where the requirement  that the two photons are in the same hemisphere is imposed, 
i.e. $\cos(\theta_{\gamma\gamma}^*) > 0$. Making the boost and applying this cut can be seen to increase the significance of the radiation zero, as shown by the dashed curve in figure~\ref{fig:r0}, at the expense of a significant loss of signal events.

It is worth noting that the radiation zero is sensitive to the AQGCs, which act to fill in the dip. However, the two-photon invariant mass was found to be a more sensitive distribution and the radiation zero will anyway be filled in by many other effects, such as background events and next to leading order contributions.

%%%%%%%%%%%%%%%%%%%%%%%%%%%%%%%%%%%%%%%%%%%%%%%%%%%%%%%%%%%%%%
%%%%%%%%%%%%%%%%%%%%%%%%%%%%%%%%%%%%%%%%%%%%%%%%%%%%%%%%%%%%%%
\section{Summary and Conclusions}

Probing the quartic gauge boson couplings represents  an important test of the non-Abelian structure of the Standard Model, and anomalous contributions to these couplings may indicate the presence of new physics, possibly in the important electroweak symmetry breaking sector.
The pp$\to\mathrm{l}^{\pm}\nu\gamma\gamma$ $(\mathrm{l}=\mathrm{l},\mu)$ process of W$\gamma\gamma$ triboson production offers an interesting starting point for the study of AQGCs at the LHC. Under the most inclusive event selection, the cross section for this process is expected to be 8.6~fb, which will yield around 260 events with an integrated luminosity of 30~fb$^{-1}$. 

After testing various distributions with a binned maximum likelihood fit, we suggest that the two-photon invariant mass will offer the best sensitivity to the anomalous coupling parameters $\beta_0$ and $\beta_{\mathrm{c}}$ associated with the WW$\gamma\gamma$ vertex. The experimental sensitivity to the anomalous couplings reaches into the region of the $(M_{\gamma\gamma}, \beta)$ plane where the effective Lagrangian theory breaks down, and a cutoff must be applied to the $M_{\gamma\gamma}$ scale to ensure unitarity conservation. Beyond this cutoff, the new physics would be directly visible in other channels. Working below the cutoff, the expected limits on the bare couplings remain around two orders of magnitude tighter than those currently available from LEP.

When studying the pp$\to\mathrm{l}^{\pm}\nu\gamma\gamma$ events, it is interesting also to plot the ``signed'' distribution of  $\Delta \eta = \eta_{\gamma\gamma} - \eta_{\mathrm{l}}$ which should reveal the radiation zero present in the SM amplitude. This can be enhanced by cutting on the opening angle of the two photons in the centre of mass system, but will require more than 30~fb$^{-1}$ of data to be experimentally observable.

%To allow the events to be passed through a realistic detector simulation in future work, we have adapted the MC programme used here to write out unweighted events in the Les Houches format. In doing so, the confidence in the MC predictions has benefitted from comparison with previously published, independent work.

%%%%%%%%%%%%%%%%%%%%%%%%%%%%%%%%%%%%%%%%%%%%%%%%%%%%%%%%%%%%%%
%%%%%%%%%%%%%%%%%%%%%%%%%%%%%%%%%%%%%%%%%%%%%%%%%%%%%%%%%%%%%%
\section{Acknowledgments}

The author would like to thank Ulrich Baur, Oscar Eboli and Sergio Morais Lietti for the provision of and agreement to use the MC generators. The work could not have been done without the support of Sergio Morais Lietti in understanding the MC of the same name. Useful discussions with Mike Seymour, Thomas LeCompte and Dave Charlton are also gratefully acknowledged, as was the technical assistance of Dusan Reljic.


\end{document}